\documentclass[10pt,conference,fleqn]{IEEEtran}

\usepackage{graphicx}
\usepackage{moreverb}
\usepackage{graphicx}
\usepackage{float}
\usepackage{amssymb}
\usepackage{amsfonts}
\usepackage{multicol}
\usepackage{amsmath}
\usepackage{mathtools}
\usepackage[mathscr]{euscript}
\usepackage{booktabs}
\usepackage{makecell}
\usepackage{multirow}
\usepackage{amsbsy} 
\usepackage[hypcap=false, font=small]{caption}
\usepackage[noend,ruled,linesnumbered]{algorithm2e}
\usepackage{cite}
\usepackage{xfrac}
\usepackage{bbm}
\usepackage{subcaption}
\usepackage[printonlyused,nolist]{acronym} 
\usepackage{nccmath}
\usepackage{lipsum}

\usepackage{tikz}

\newcommand\copyrighttext{%
  \footnotesize \textcopyright~\the\year~IEEE. Personal use of this material is permitted.
  Permission from IEEE must be obtained for all other uses, in any current or future
  media, including reprinting/republishing this material for advertising or promotional
  purposes, creating new collective works, for resale or redistribution to servers or
  lists, or reuse of any copyrighted component of this work in other works.}
\newcommand\copyrightnotice{%
\begin{tikzpicture}[remember picture,overlay]
\node[anchor=north,yshift=-15pt] at (current page.north) {\fbox{\parbox{\dimexpr\textwidth-\fboxsep-\fboxrule\relax}{\copyrighttext}}};
\end{tikzpicture}
\vspace{-0.3cm}
}

\IEEEoverridecommandlockouts 

\title{Mobility-Aware Joint Service Placement and Routing in Space-Air-Ground Integrated Networks\vspace{-.5ex}}

\author{\IEEEauthorblockN{Amir Varasteh\IEEEauthorrefmark{1}, Sandra Hofmann\IEEEauthorrefmark{2}, Nemanja Deric\IEEEauthorrefmark{1}, Mu He\IEEEauthorrefmark{1}, Dominic Schupke\IEEEauthorrefmark{2}, \\ Wolfgang Kellerer\IEEEauthorrefmark{1}, and Carmen Mas Machuca\IEEEauthorrefmark{1}
\IEEEauthorblockA{\IEEEauthorrefmark{1}Chair of Communication Networks, Department of Electrical and Computer Engineering,\\
Technical University of Munich, Germany\\Email: \{amir.varasteh, nemanja.deric, mu.he, wolfgang.kellerer, cmas\}@tum.de}
\IEEEauthorblockA{\IEEEauthorrefmark{2}Airbus, Munich, Germany\\Email: \{sandra.s.hofmann, dominic.schupke\}@airbus.com\vspace{-3ex}}
}}

\begin{document}
\maketitle
\copyrightnotice 

\begin{abstract}
People desire to be \textit{connected}, no matter where they are. Recently, providing Internet access to on-board passengers has received a lot of attention from both industry and academia. However, in order to guarantee an acceptable \ac{QoS} for the passenger services with low incurred cost, the path to route the services, as well as the datacenter~(DC) to deploy the services should be carefully determined. This problem is challenging, due to different types of \ac{A2G} connections, i.e., satellites and \ac{DA2G} links. These A2G connection types differ in terms of cost, bandwidth, and latency. Furthermore, due to the flights' movements, it is important to consider adapting the service location accordingly. In this work, we formulate two Mixed Integer Linear Programs~(MILPs) for the problem of Joint Service Placement and Routing (JSPR): \textit{i)}~Static~(S-JSPR), and \textit{ii)}~Mobility-Aware~(MA-JSPR) in Space-Air-Ground Integrated Networks~(SAGIN), with the objective of minimizing the total cost. We compare S-JSPR and MA-JSPR using comprehensive evaluations in a realistic European-based SAGIN. The obtained results show that the MA-JSPR model, by considering the future flight positions and using a service migration control, reduces the long-term total cost notably. Also, we show S-JSPR benefits from a low time-complexity and it achieves lower end-to-end delays comparing to MA-JSPR model.
\end{abstract}
\vspace{-0.2cm}
\maketitle
\section{Introduction}\label{sec:intro}
\vspace{-.05cm}
The Internet connectivity to the passengers of flights is being offered by more and more airlines. It has been shown that passengers consider connectivity as the third important parameter when choosing a specific flight~\cite{wifiimportance}. Different services can be delivered to passengers during the flight, such as Voice-over-IP (VoIP) calls, video streaming, and web browsing. In addition to passengers, flight's operational services need to communicate to the ground, such as sensor information for monitoring the status of the flight. Hence, the services between flights and ground locations (usually at datacenters~(DCs)) are increasing in terms of number of services as well as \ac{QoS} requirements. With the growing number of connected flights as well as connected passengers, it becomes challenging to guarantee the \ac{QoS} requirements of each service.

\begin{figure}[t]
\centering
\includegraphics[width=8.7cm, height=6.67cm]{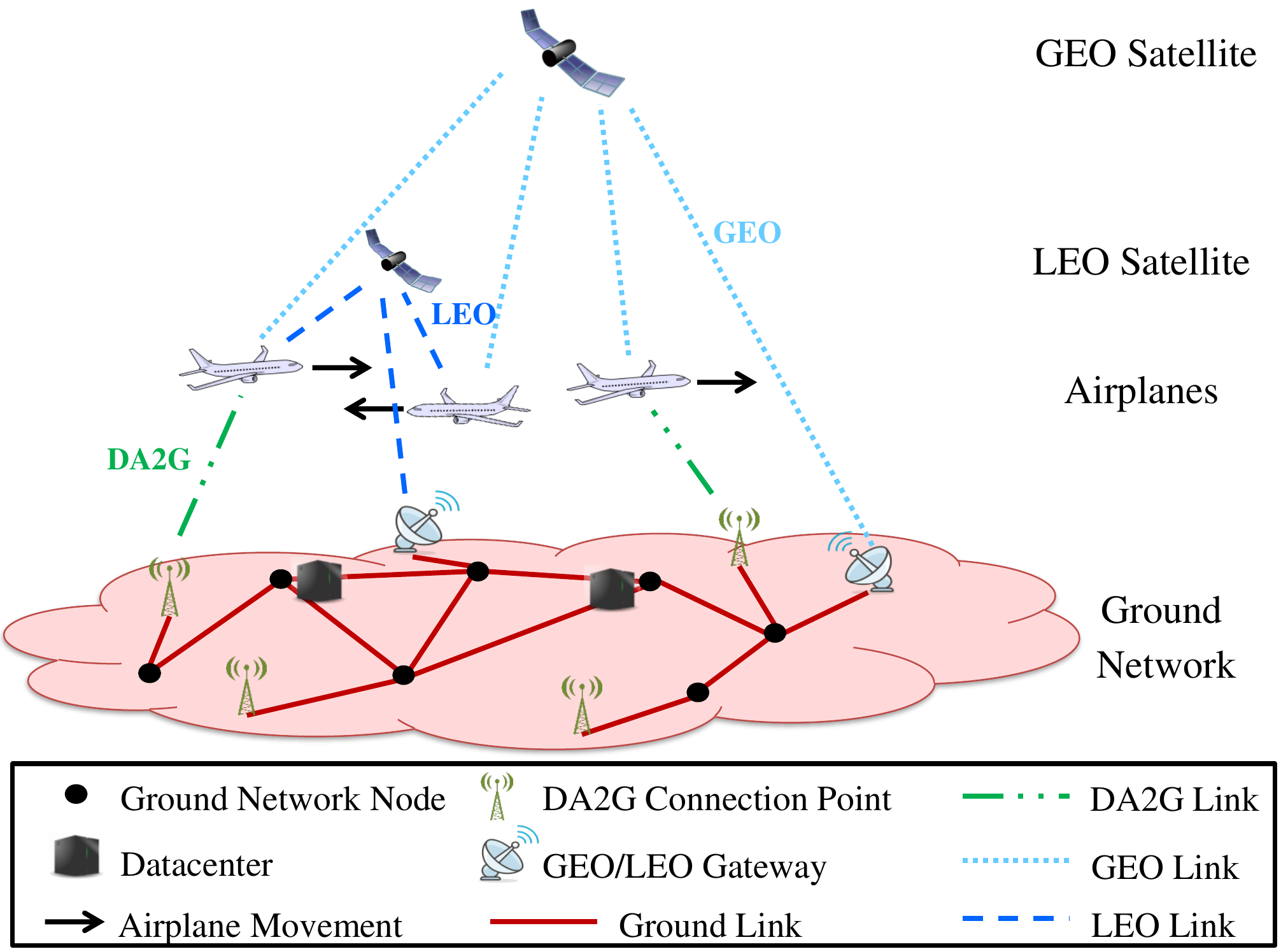}
\caption{A high level view of the space-air-ground integrated network}\label{fig:scenario}
\end{figure}

As depicted in Fig.~\ref{fig:scenario}, a flight has different Air-To-Ground~(A2G) alternatives to connect to a particular DC in the ground network: \textit{i)} Satellites, \textit{ii)} Direct Air-To-Ground~(\ac{DA2G}) links. In this paper, we discuss two types of satellites: \ac{GEO} and \ac{LEO}. GEO satellites are designed to have an orbital period equal to the Earth's rotation period. However, due to the high orbit altitude of these satellites, their \ac{RTT} is over $600~ms$~\cite{qu2017leo}, being too large for some services. Besides, OneWeb, Telesat, and SpaceX have been developing LEO satellite constellations, which are located at lower altitude (less than $2,000$~km)~\cite{qu2017leo}. Thus, LEO satellites are able to offer network bandwidth with low latency (the average \ac{RTT} is $50~ms$~\cite{leodelayandposition}). The satellites are connected to their gateways, located in the ground network. 

The second option to connect a flight to the ground network is \ac{DA2G}, which establishes a direct connection from a flight to the DA2G base station. A DA2G link produces around $10~ms$ of \ac{RTT}~\cite{dinc2017multi}. It has to be mentioned, that the connection to DA2G base stations change due to the mobility of flights over time. Though, different A2G connection options differ in terms of provided bandwidth, delay, and cost.

On the other hand, a wise mapping of service instances (e.g., Virtual Machines~(VMs)) to proper DCs can improve QoS levels and reduce the long-term incurred costs~\cite{gu2015optimal,varasteh2017server}. Additionally, by considering the mobility of flights, inter-DC service migrations can improve the service placement accordingly. Thus, different cost parameters can contribute to the total cost of providing Internet services to a flight: \textit{i)} routing cost from the flight to the ground DC, \textit{ii)} service instance deployment cost, and \textit{iii)} the cost of service migration between DCs. Therefore, in such dynamic and heterogeneous environment, providing cost-efficient and QoS-guaranteed services to flights is still a quite challenging problem to overcome, which to the best of our knowledge, has not been addressed in the state-of-the-art.

This paper tackles the problem of Joint Service Placement and Routing~(JSPR) for on-board passenger services in a European-based SAGIN by minimizing the total cost. We formulated the JSPR problem in two cases: \textit{i)}~Static: it solves the JSPR for a particular time-slot (e.g., position of the flight, DC status), \textit{ii)}~Mobility-Aware (i.e., dynamic): This approach takes future time-slots into account and hence, the flight's mobility to solve the JSPR problem. We also introduce and model a service instance migration control approach to improve the solution in terms of total cost.

The remainder of the paper is structured as follows: In Section~\ref{sec:relatedwork}, we highlight the related work. Section~\ref{sec:problem} presents the system model and problem formulation. Further, in Section~\ref{sec:performance}, we evaluate and compare the performance of S-JSPR and MA-JSPR optimization approaches in a European-based SAGIN to identify their strengths and weaknesses. Finally, Section~\ref{sec:conclusion} concludes the paper.
\vspace{-.12cm}
\section{Related Work}\label{sec:relatedwork}
\vspace{-.05cm}
In a SAGIN architecture space, air and ground network segments are integrated~\cite{evans2005integration}. Many research works exist in the the area of SAGIN. Some studies have focused on resource allocation~\cite{cao2018gateway}, mobility management~\cite{xu2016study}, reliability~\cite{cao2018optimal}, and energy-efficiency~\cite{shi2018inter} in these networks. Further, with the recent advances in Software-Defined Networking~(SDN), Zhang~et.~al. in~\cite{zhang2017software} present an SDN-based vehicular network utilizing SAGIN. Furthermore, in~\cite{wu2018dynamic,arledpaper}, the SDN controller placement problem in these integrated networks is explored. Comprehensive surveys summarize studies on the SAGIN aforementioned challenges, available in~\cite{liu2018space}.

In addition to SAGIN, looking from the modeling aspect, some studies in the area of Vehicular Cloud Computing~(VCC)~\cite{yao2015migrate,aissioui2018enabling} and Mobile Edge Computing~(MEC)~\cite{taleb2015user,bahreini2017efficient,farris2017optimizing,wang2017online,wang2017dynamic} 
have developed mobility-aware resource management approaches. On the one hand, in the area of VCC, authors in~\cite{aissioui2018enabling} present an SDN-based platform to guarantee QoS requirements for 5G-enabled automotive systems. In their approach, while the vehicles are moving, their QoS level is being continuously monitored. In case of losing certain QoS levels, the service migration is triggered for QoS improvement. Moreover, Yao~et.~al.~\cite{yao2015migrate} assign a VM per vehicle to deliver the required services, such as in-vehicle multimedia entertainment and vehicular social networking. By minimizing the network costs, they determine the necessity and the place to migrate the VMs during vehicle movement. However, they do not determine the routing from vehicles to their respective VM instances. Also, in contrast to us, they solve the problem for the next time-slot, instead of the full time horizon. 

On the other hand, in the MEC area, authors in~\cite{taleb2015user,farris2017optimizing} propose an approach to dynamically place network functions on MEC servers available in mobile base stations, according to the handover probability of users for the next time-slot. They formulate two optimization models with the objective of minimizing network function migrations and communication cost (i.e., QoS/QoE) between users (UEs) and network functions. In another work, Bahreini~et.~al.~\cite{bahreini2017efficient} formulate an optimization model to map application graphs to network graphs according to the user's location. Their objective is to minimize the total cost of running and migrating applications, and communication cost between users and applications. Finally, a dynamic service placement approach is proposed by Wang~et.~al.~\cite{wang2017dynamic}. They minimize the placement, routing, and migration costs based on a specific look ahead window. However, they do not determine the routing between users and instances. Also, they do not guarantee the QoS of delivered services. A comparison of our approach with others is summarized in Table~\ref{table1}. Notably, \textit{shared instance} column in Table~\ref{table1} indicates whether a service instance is shared among several flights at the same time.

To the best of our knowledge, this is the first paper that formulates and evaluates the JSPR problem in a European-based SAGIN, by minimizing the total cost. Further, this work provides a service migration control to adapt service placement and routing according to the mobility of flights. We also guarantee to meet the end-to-end delay, required by passenger services.

\begin{table}[t]
\centering
\setlength\belowcaptionskip{7pt}
\caption{Comparison of the related work with our approach, all considering mobility-awareness}
\label{table1}
\resizebox{\columnwidth}{!}{
\begin{tabular}{c||ccccc}
\toprule
\textbf{Works} & \textbf{Routing}  & \textbf{Placement} & \textbf{Service Migration} &  \textbf{QoS} & \textbf{Shared Instance} \\ \hline\hline
\cite{farris2017optimizing,yao2015migrate} & $\times$ (predefined) & $\checkmark$ & $\checkmark$ & $\times$ & $\times$ \\ \hline
\cite{taleb2015user,wang2017online,aissioui2018enabling} & $\times$ & $\checkmark$ & $\checkmark$ & $\checkmark$ & $\times$ \\ \hline
\cite{bahreini2017efficient,wang2017dynamic} & $\times$ & $\checkmark$ & $\checkmark$ & $\times$ & $\times$ \\ \hline
\textbf{Our work} & $\checkmark$ & $\checkmark$ & $\checkmark$ & $\checkmark$ & $\checkmark$ \\ 
\bottomrule
\end{tabular}}
\end{table}

\vspace{-.1cm}
\section{System Model and Problem Formulation}\label{sec:problem}
\vspace{-0.03cm}
Before presenting the JSPR problem formulation cases, let us introduce the system model. 
\vspace{-.07cm}
\subsection{System Model}
\vspace{-.04cm}
Let us define a finite time horizon $\mathcal{T}=\{t_0,t_1,...,|\mathcal{T}|\}$ comprising $|\mathcal{T}|$ time-slots, each with a duration of $\omega$ minutes. The SAGIN considered in this work consists of three segments:

\textit{1) Ground Network}: We denote the set of nodes and links of the ground network as $N_G$ and $L_G$, respectively. Further, we consider a set of DCs as $N_{DC} \subset N_G$, which can host service instances. We note that service instances can be deployed as VMs in DCs. We consider each $j \in N_{DC}$ with a total resource capacity of $Cap^{DC}_{j}$ (e.g., in terms of CPU cores).

\textit{2) DA2G Nodes/Links}: Flights can communicate to the ground via DA2G links and base stations. We define $N_{DA2G}$ as the set of DA2G base stations in the ground network, such that each $N_{DA2G}$ node is connected to the closest network node $N_G$. Also, we denote the set of links that connect flights to the closest DA2G base station as $L_{DA2G}$.

\textit{3) Satellite Nodes/Links}: We denote the set of satellite nodes and links by $N_{SAT}$ and $L_{SAT}$, respectively. The satellite nodes relay the flight's traffic to the satellite gateways located in the ground network, defined as $N_{SG}$~\cite{leodelayandposition}.

Thus, SAGIN is defined as a bidirectional connected graph $\mathcal{G}=(\mathcal{N},\mathcal{L})$, where $\mathcal{N}=N_{SAT} \cup N_{DA2G} \cup N_G \cup N_A$ and $\mathcal{L} =L_{SAT} \cup L_{DA2G} \cup L_G$ are the set of all nodes and links of $\mathcal{G}$, respectively. We note that, each link $(u,v) \in \mathcal{L}$ between nodes $u,v \in \mathcal{N}$ is characterized by a constant cost value $Cost^L_{uv}$, propagation delay $D_{uv}$, and capacity $B_{uv}$.

We consider $\mathcal{K}$ as the set of all service types offered to the passengers. The service requests triggered by a passenger is denoted as $a \in A$ and is characterized by following 4-tuple: $(src_a, k_a, b_a, d_a) $, where $src_a \in N_A$ is the flight node of the passenger, $k_a \in \mathcal{K}$ is the service type of request $a$, $b_a$ is the required bandwidth, and $d_a$ is the maximum allowed end-to-end delay of the service request $a$.

Let us denote $\mathcal{F}$ as the set of flights considered in the problem (e.g., flights flying over Europe during the considered time horizon $\mathcal{T}$). Each flight $F_i \in \mathcal{F}$ has a different flight route and thus, it is characterized by the different locations (in terms of Latitude~(lat) and Longitude~(lon)) and the duration of the flight. We define $\tau_i = \mid {F}_i \mid-1$ as the number of time-slots in which the $i^{th}$ flight is flying. Thus, the duration of the flight $i$ can be calculated as $\tau_i \times \omega$. The flight locations at any time-slot are considered as nodes in graph $\mathcal{G}$, which are denoted as $N_A$ where $\mid N_A \mid = \sum_{i=1}^{\mid \mathcal{F} \mid} \mid F_i\mid$. As an example, suppose there is only one flight as $\mathcal{F}=\{F_1\}$ where $\tau_1 = 2$, and $F_1 = \{(lat_1,lon_1), (lat_2,lon_2), (lat_3,lon_3)\}$. In this example, $N_A$ would be the set of $\mid F_1 \mid= 3$ flight nodes, in the respective positions at each time-slot. Considering the flight duration $\omega = 30$ minutes, the duration of the $i^{th}$flight would be $\tau_i \times 30 = 60$ minutes. We now introduce the two JSPR formulations:
\vspace{-.1cm}
\subsection{Static JSPR~(S-JSPR) Formulation}\label{sec:s-jspr}
\vspace{-.05cm}
The static case is defined as follows: given a SAGIN and a set of service requests in each time-slot $t \in \mathcal{T}$, place the required service instances in the DCs and find the route to the passengers, such that the network and service instance deployment costs are minimized and all the service requirements (bandwidth and maximum delay) are guaranteed. Inspired by~\cite{papadimitriou2018mixed}, we formulate S-JSPR by integrating the classical multi-commodity flow and capacitated facility location-routing problems into a MILP optimization. Therefore, we define four types of decision variables: \textit{i)} $x_{aj} \in \{0,1\}$: if service request $a \in A$ is assigned to DC $j \in N_{DC}$. \textit{ii)} $l^{uv}_{aj}\in \{0,1\}$: if link $(u,v) \in \mathcal{L}$ is used to route the traffic for service request $a \in A$, assigned to DC $j \in N_{DC}$. \textit{iii)} $n_{jk}\geq 0$: The number of service instances deployed in DC $j \in N_{DC}$ to provide service type $k \in \mathcal{K}$. \textit{iv)}~$y_{uv}\in \{0,1\}$: if link $(u,v) \in \mathcal{L}$ is active.

Therefore, we present the S-JSPR optimization formulation as follows:
{\small{
\begin{flalign}
    & \text{Min.} \hspace{.15cm} \big({\sum_{j\in N_{DC}} \sum_{k \in \mathcal{K}} Cost^{IN}_k n_{jk}} + \sum_{(u,v) \in \mathcal{L}} Cost^L_{uv} y_{uv}\big) \label{eq:s-jspr},\\
    & \text{s.t.}\hspace{.2cm}\sum_{j \in N_{DC}} x_{aj} =1,\forall a \in A \tag{1.1} \label{eq1.1},\\
    &x_{aj} \leq n_{jk_a},\forall a\in A, \forall j \in N_{DC} \tag{1.2}\label{eq1.2},\\
    &\sum_{k \in \mathcal{K}} Size_k^{IN}   n_{jk} \leq Cap^{DC}_j, \forall j \in N_{DC} \tag{1.3}\label{eq1.3}, 
\end{flalign}
\begin{flalign}
    &\sum_{a\in A} b_a   x_{aj} \leq Cap^{IN}   n_{j k_a} ,\forall j \in N_{DC} \tag{1.4}\label{eq1.4},\\
    &\sum_{a \in A} \sum_{j \in N_{DC}} l^{uv}_{aj}   b_a \leq B_{uv},\forall (u,v) \in \mathcal{L}, \tag{1.5}\label{eq1.5},\\
    &\sum_{(u,v) \in \mathcal{L}} \sum_{j \in N_{DC}} l^{uv}_{aj}   D_{uv} \leq d_a,\forall a \in A \tag{1.6}\label{eq1.6},\\
    &\sum_{v \in \Psi^+(u)} \sum_{j \in N_{DC}} l^{src_av}_{aj} = 1,\forall a \in A \tag{1.7}\label{eq1.7},\\
    &\sum_{v \in \Psi^+(u)} \sum_{j \in N_{DC}} l^{vu}_{aj} - \sum_{v \in \Psi^-(u)} \sum_{j \in N_{DC}} l^{uv}_{aj} =\nonumber \\
    & \begin{cases}
    0&,\forall a \in A, \forall u \in \{ \mathcal{N} \setminus N_{DC}\}, u \neq src_a \\
    x_{au}&, \forall a \in A, \forall u \in N_{DC} \nonumber
  \end{cases} ,\tag{1.8} \label{eq1.8}\\
    &l^{uv}_{aj} \leq x_{aj}, \forall (u,v)\in \mathcal{L}, \forall a \in A, \forall j \in N_{DC} \tag{1.9}\label{eq1.9},\\
    &y_{uv} \leq \sum_{a \in A} \sum_{j \in N_{DC}} l^{uv}_{aj} , \forall (u,v) \in \mathcal{L} \tag{1.10}\label{eq1.10},\\
    & \rho y_{uv} \geq \sum_{a \in A} \sum_{j \in N_{DC}} l^{uv}_{aj} , \forall (u,v) \in \mathcal{L} \label{eq1.11}\tag{1.11}.
\end{flalign}}}
\noindent Eq.~(\ref{eq:s-jspr}) forms the S-JSPR objective function, aiming at minimizing the sum of service instance deployment costs and routing costs (i.e., used links), where $Cost^{IN}_k$ is the deployment cost of a single service instance for service type $k \in \mathcal{K}$. Also, $Cost^L_{uv}$ is the cost of using link $(u,v) \in \mathcal{L}$. Constraint~(\ref{eq1.1}) ensures that each service request is assigned to one DC. Constraint~(\ref{eq1.2}) states that a service request of type $k$ can be assigned to a DC, if and only if there is a service instance with type $k$ is deployed on the DC $j$. Constraint~(\ref{eq1.3}) expresses the DC capacity limit (e.g., in terms of CPU cores), where $Size_k^{IN}$ is the resource requirement of the service instance type $k$. Also, $Cap^{DC}_j$ is the maximum capacity of DC node $j \in N_{DC}$. Constraint~(\ref{eq1.4}) guarantees that the amount of assigned requests to an instance does not exceed the maximum capacity of the instance, in which $b_a$ is the amount of bandwidth requested by service request $a$. Constraint~(\ref{eq1.5}) expresses that the sum of assigned traffic volume to a link cannot exceed its capacity. Constraint~(\ref{eq1.6}) guarantees the QoS of each service request (in terms of end-to-end delay). Constraint~(\ref{eq1.7}) generates the traffic from each request source, where $\Psi^+(u)$ is the set of outgoing links from node $u$. Constraint~(\ref{eq1.8}) represents the flow conservation rule, where $\Psi^-(u)$ is the set of incoming links to node $u$. In detail, this constraint states that the traffic should keep flowing, except when a node is a DC node serving the service request. Constraint~(\ref{eq1.9}) states that traffic routing must be done for the service requests that are actually assigned to a DC. Constraints~(\ref{eq1.10})-(\ref{eq1.11}) activate a link, if any service request is routing through it ($\rho\gg0$).
\vspace{-.1cm}
\subsection{Mobility-Aware JSPR~(MA-JSPR) Formulation}\label{sec:ma-jspr}
\vspace{-.05cm}
Out MA-JSPR model considers mobility of flights in order to optimize the long-term total cost over time horizon $\mathcal{T}$. It also considers service instance migration control between DCs. Therefore, MA-JSPR adds a new dimension to S-JSPR: \textit{time}. In this case, the questions addressed in this problem are: \textit{i)} Considering the mobility of flights (future positions), how many service instances, and on which DC should be placed to serve the passenger service requests over time horizon?~(\textit{Placement}) \textit{ii)} How to reach from the passenger's flights to each service instance during the flight given ($\mathcal{T}$)?~(\textit{Routing}) \textit{iii)} At which point in time, a service instance should be reallocated to another DC?~(\textit{Service migration control}). 

MA-JSPR addresses these questions by minimizing the total cost over the whole time horizon. In MA-JSPR, there is an additional cost associated to the service instance migration (referred as migration cost $Cost^{mig}$), which is added to the service instance deployment $Cost^{IN}$, and routing costs $Cost^{L}$ of S-JSPR. To formulate MA-JSPR, we define $A^t$ as the set of service requests generated by passengers on all flights at time-slot $t$. We extend the variables defined in S-JSPR to include a per time-slot index $t \in \mathcal{T}$. Therefore, by using the same definition as in S-JSPR model, we convert the variables of S-JSPR to multi-period variables as $x_{ij} \rightarrow x_{ij}^t$, $l^{uv}_{aj} \rightarrow l^{uvt}_{aj}$, $n_{jk} \rightarrow n_{jk}^t$, and $y_{uv} \rightarrow y_{uv}^t$. Additionally, we define a new variable $m^t_k \geq 0$ to include the service instance migration control. $m^t_k \geq 0$ represents the number of service migrations for service type $k \in \mathcal{K}$ at the time-slot $t \in \mathcal{T}$. 

Considering the above explanations, we present the MILP formulation of MA-JSPR as follows:

{\small{\begin{flalign}
    &\text{Min.} \hspace{.15cm} \big(\sum_{t \in \mathcal{T}}\sum_{j\in N_{DC}} \sum_{k \in \mathcal{K}} Cost^{IN}_k n^t_{jk} +\nonumber \\
    & \hspace{.75cm}\sum_{t \in \mathcal{T}}\sum_{(u,v) \in \mathcal{L}} Cost^L_{uv} y^t_{uv} +\sum_{t \in \mathcal{T}} \sum_{k \in \mathcal{K}} Cost^{mig}_k  m^t_k\big) \label{eq:ma-jspr},\\
    & \text{s.t.}\hspace{.2cm} \sum_{j \in N_{DC}} x_{aj}^t =1,\forall t \in \mathcal{T}, \forall a \in A^t \tag{2.1}\label{eq2.1}, \\
    &x_{aj}^t \leq n_{j k_a}^t,\forall a\in A^t, \forall t \in \mathcal{T},\forall j \in N_{DC} \tag{2.2} \label{eq2.2}, \\
    &\sum_{k \in \mathcal{K}} Size_k^{IN}   n_{jk}^t \leq Cap^{DC}_j, \forall t \in \mathcal{T}, \forall j \in N_{DC} \tag{2.3} \label{eq2.3},\\
    &\sum_{a\in A^t} b_a   x_{aj}^t \leq Cap^{IN} n_{j k_a}^t ,\forall t \in \mathcal{T} ,\forall j \in N_{DC} \tag{2.4}\label{eq2.4},\\
    &\sum_{a \in A^t} \sum_{j \in N_{DC}} l^{uvt}_{aj}   b_a \leq B_{uv}, \forall t \in \mathcal{T},\forall (u,v) \in \mathcal{L} \tag{2.5}\label{eq2.5},\\
    &\sum_{(u,v) \in \mathcal{L}} \sum_{j \in N_{DC}} l^{uvt}_{aj}   D_{uv} \leq d_a, \forall t \in \mathcal{T},\forall a \in A^t \tag{2.6}\label{eq2.6},\\
    & \sum_{v \in \Psi^+(u)} \sum_{j \in N_{DC}} l^{src_avt}_{aj} = 1,\forall t \in \mathcal{T},\forall a \in A^t \tag{2.7}\label{eq2.7},\\
    &\sum_{v \in \Psi^+(u)} \sum_{j \in N_{DC}} l^{vut}_{aj} - \sum_{v \in \Psi^-(u)} \sum_{j \in N_{DC}} l^{uvt}_{aj}= \tag{2.8}\label{eq2.8} \\
& \begin{cases}
0&,\forall t\in \mathcal{T}, \forall a \in A^t, \forall u \in \{ \mathcal{N} \setminus N_{DC}\}, u \neq src_a \nonumber \\
x_{au}^t&, \forall t \in \mathcal{T}, \forall a \in A^t, \forall u \in N_{DC} \nonumber
\end{cases},\\
    &y_{uv}^t \leq  \sum_{a \in A^t} \sum_{j \in N_{DC}} l^{uvt}_{aj} , \forall t \in \mathcal{T},\forall (u,v) \in \mathcal{L} \tag{2.9}\label{eq2.9},\\
    &l^{uvt}_{aj} \leq x_{aj}^t, \forall (u,v)\in \mathcal{L}, \forall t \in \mathcal{T}, \forall a \in A^t, \forall j \in N_{DC} \tag{2.10} \label{eq2.10},
\end{flalign}
\begin{flalign}
    &\rho y_{uv}^t \geq \sum_{a \in A^t} \sum_{j \in N_{DC}} l^{uvt}_{aj} , \forall t \in \mathcal{T}, \forall (u,v) \in \mathcal{L} \tag{2.11}\label{eq2.11}.
\end{flalign}}}
\noindent Eq.~(\ref{eq:ma-jspr}) represents the objective function for MA-JSPR, where $Cost^{mig}_k$ (third term) is defined as the migration cost for service type $k$. Constraints~(\ref{eq2.1})-(\ref{eq2.11}) express the Constraints (\ref{eq1.1})-(\ref{eq1.11}) in S-JSPR model, extended by time dimension, respectively. For instance, constraint~(\ref{eq2.5}) makes sure that the assigned bandwidth to each link $(u,v) \in \mathcal{L}$ does not exceed the total bandwidth of the link at each time-slot $t \in \mathcal{T}$. 

In order to compute the number of migrations of service type $k \in \mathcal{K}$ at the time-slot $t \in \mathcal{T}$, we propose Eq.~\ref{eq:mig}:
{\small{\begin{align}
    m^t_k =& \sum_{j \in N_{DC}} [n_{jk}^t - n_{jk}^{t-1}]^+ - [\sum_{j \in N_{DC}} n_{jk}^{t} - \sum_{j\in N_{DC}} n_{jk}^{t-1} ]^+ \nonumber\\
    & ,\forall k \in \mathcal{K}, \forall t \in \mathcal{T}^+ \label{eq:mig},
\end{align}}}
\noindent where $[a]^+$ is defined as $max\{0,a\}$. Let us note that $m^0_k=0$, since service migration cannot happen at  $t=0$. Consider the example in Fig.~\ref{fig:migration_example} with two services $\mathcal{K}=\{1,2\}$ and three DCs $A$, $B$, and $C$. Let us consider that at time-slot $t_0$, one service instance of $k=1$ is required at DC $A$, and two instances of $k=2$ are required at DCs $B$, and $C$. However, at next time-slot, due to some new flight locations, two service instances of $k=1$ are required at B and C, whereas only one service instance for $k=2$ is required at DC $A$. This new service instance allocation requires a migration of service $k=1$ from A to B, a second migration of service $k=2$ from B to A, and a service instance creation and deletion at C for service types $k=2$ and $k=1$, respectively. Applying Eq.~\ref{eq:mig} to calculate the number of service migrations for each service type $k$ ($m_t^k$), we obtain:
\begin{figure}[]
\centering
\includegraphics[width=.9\linewidth]{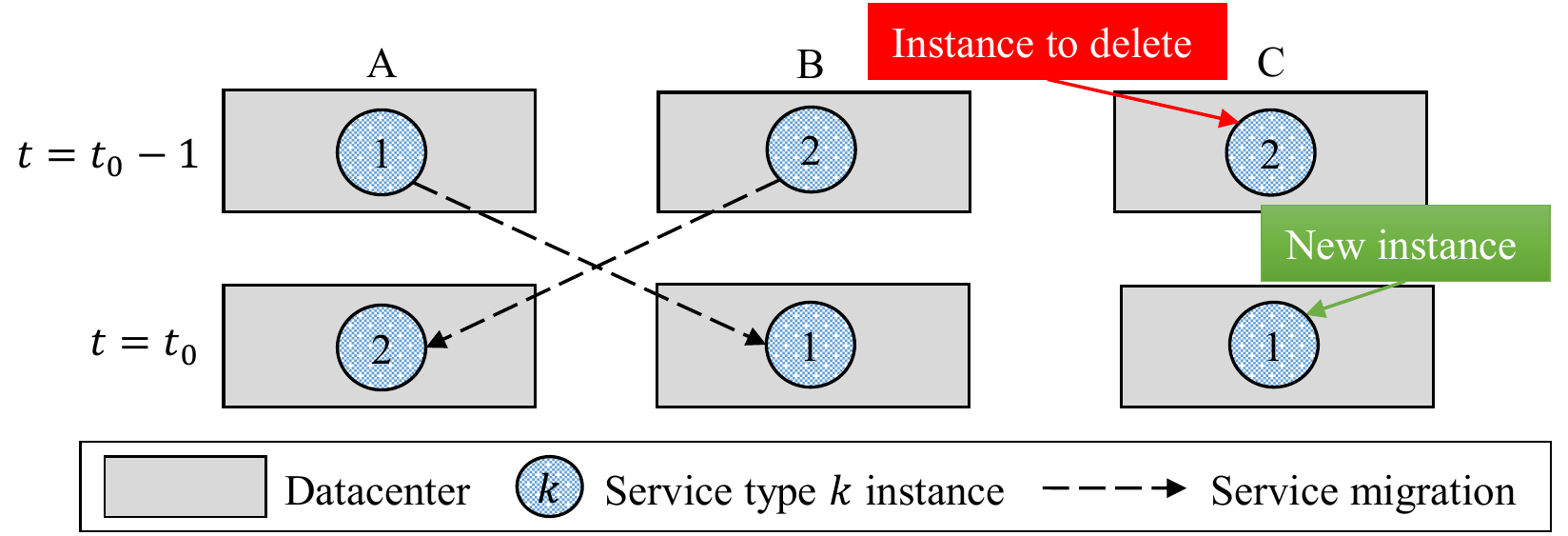}
\caption{Example of the service migration computation for three DCs A, B, and C, when two service migrations, one service deletion, and one creation is performing.}
\label{fig:migration_example}
\end{figure}

{\small{\begin{align}
(k=1): & m_1^{t_0} = [(0-1)^+ + (1-0)^+ + (1-0)^+]  \nonumber \\
            & - [(2-1)^+] = 1, \nonumber\\
(k=2): &m_2^{t_0} = [(1-0)^+ + (0-1)^+ + (0-1)^+]  \nonumber \\
            & - [(1-2)^+] = 1 .\nonumber
\end{align}}}
\noindent Thus, the total number of service instance migrations for all $k \in \mathcal{K}$ is given by $m_1^{t_0} + m_2^{t_0} = 2$.
\section{Performance Evaluation}\label{sec:performance}
In this section, we firstly introduce our scenario and input parameters. Then, we compare the performance of S-JSPR and MA-JSPR models, in terms of total cost, number of service migrations, average end-to-end delay, and runtime.

\textbf{\textit{A. Scenario:}} Our study focuses on a European-based SAGIN. For the ground network topology, we extend the PAN network topology~\cite{giorgetti2009label} to cover DA2G base stations over Europe. We set the bandwidth for $L_{G}$ to $2~Gbps$ and their delays were determined according to the length of the optical fiber transmissions. Further, to be able to serve any flight over Europe, distributed DC locations have been selected~\cite{dclocation}: Athens (Greece), Helsinki (Finland), Liverpool (England), Strasbourg (France), Madrid (Spain), and Lviv (Ukraine). The locations of the DA2G base stations are estimated using~\cite{da2gposition} which are distributed all over Europe ($\mid N_{DA2G} \mid = 295$). Also, the bandwidth and delay of $L_{DA2G}$ is considered as $75~Mbps$ and $10~ms$~\cite{dinc2017multi}, respectively. 

According to the results presented in~\cite{arledpaper}, we observe that the number of LEO satellites moving over Europe is rather low ($5$ out of the $72$ LEO satellites in Iridium constellation). Compared to GEO, utilizing LEO satellites introduces lower latency, which is more suitable for our use case. Therefore, in this paper, we only consider the LEO constellation for modeling the problem. For simplification purposes, we apply an abstraction model, where the group of LEO satellites occupying Europe is represented as a single satellite node denoted as set $N_{SAT} \subset \mathcal{N}$, where $\mid N_{SAT}\mid =1$. Also, the set of links that connects each flight to LEO node and from LEO to the satellite gateway is denoted as $L_{SAT}$. We set the bandwidth of $L_{SAT}$ to $50~Mbps$~\cite{leodelayandposition}. Moreover, to avoid unrealistic assumptions, we set the latency of $L_{SAT}$ to the worst-case achievable latency $50~ms$ (according to LEO latency calculations provided in~\cite{mcmahon2005measuring}). Also, two satellite gateways $N_{GS}$ have been considered in Florence (Italy) and Patras (Greece)~\cite{leodelayandposition}. The set of considered flights has been exported from FlightRadar24 live air traffic for $24$ hours on $9.11.2017$. In our experiments, assuming $\omega = 30$ minutes, we consider two set of flights with different duration: \textit{i)} \textit{Short} flight $\mathcal{F}_s$, such that $\tau_s = 3$, $\forall F \in \mathcal{F}_s$; and \textit{ii)} \textit{Long} flight $\mathcal{F}_l$, where $\tau_l = 7$, $\forall F\in \mathcal{F}_l$. Without loss of generality, we set the visibility distance from each flight to the closest DA2G base station as 350~km as the absolute geometrical maximum~\cite{hoffmann2015routing}, although in reality it can vary based on e.g., antenna type and weather conditions. 

In order to reduce the time complexity of the optimization, two aggregated service requests per flight per time-slot have been considered. These two service types are defined as: \textit{i)} Video streaming with bandwidth and end-to-end delay requirement of $1.5~Mbp$s~\cite{videobw} and $300~ms$~\cite{videodelay}, respectively, \textit{ii)} VoIP with $64~Kbps$ bandwidth and $100~ms$~\cite{varasteh2018power} delay requirement. We assumed the flight types as Airbus $A320$ with $150$ passengers~\cite{airbusA320}, where $20\%$ of the passengers use the aforementioned network services~\cite{alliance20155g}. We assume these requests are divided between our two service types equally.

\begin{figure}[t]
\centering
\includegraphics[width=.6\linewidth]{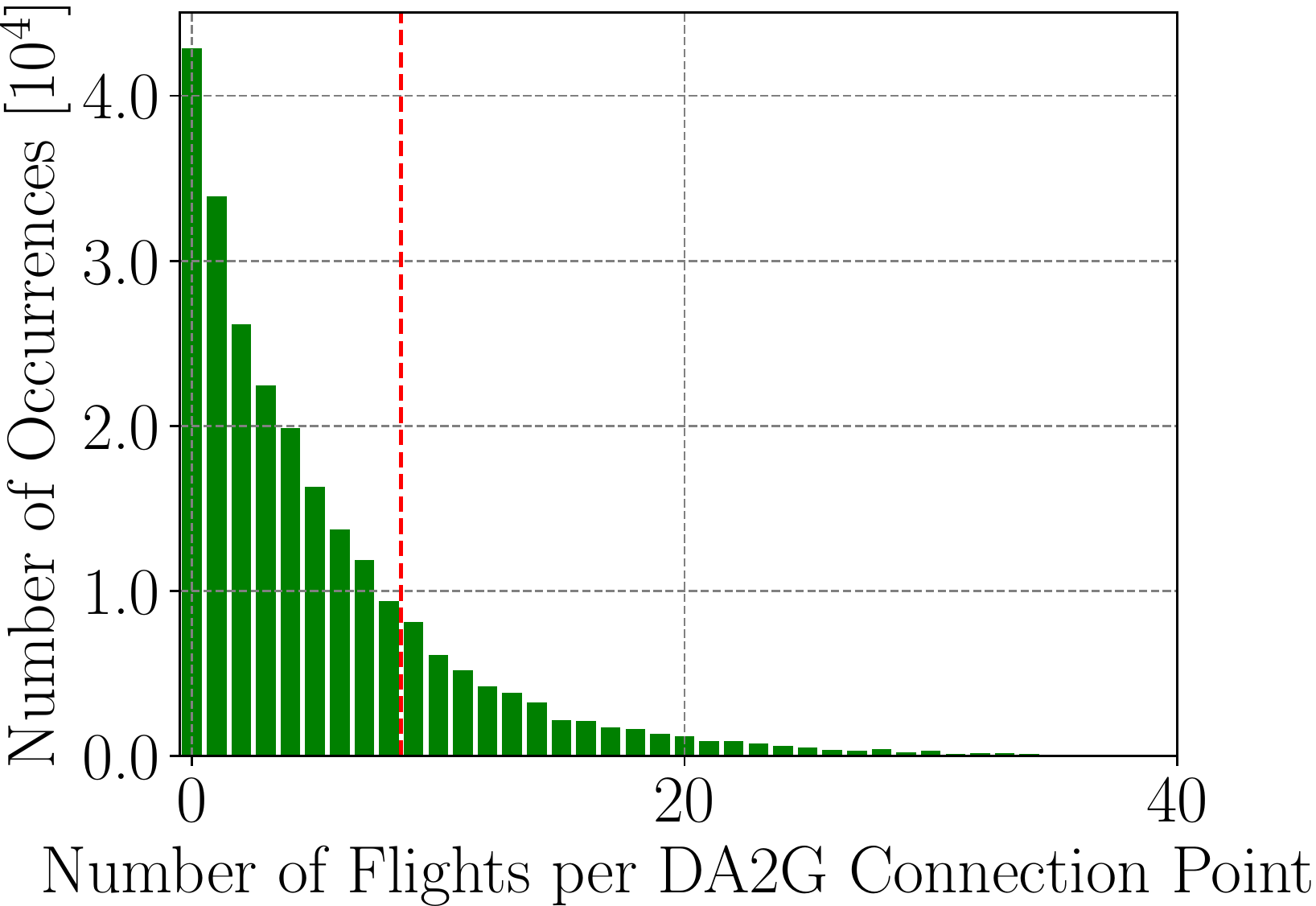}
\caption{Number of flights in a DA2G base station coverage range, based on the real data of flights for 24 hours over Europe.}
\label{fig:da2gcongestion}
\end{figure}

First of all, based on our data, let us consider the number of flights associated to each DA2G base station in order to calculate the probability that a DA2G link is congested. Fig.~\ref{fig:da2gcongestion} shows the histogram summarizing all the flights over Europe in $24$ hours. 
According to our flight data, the DA2G capacity, and the services required by a flight, every DA2G base station can serve simultaneously only around $9$ flights (see the vertical line in Fig.~\ref{fig:da2gcongestion}). Based on this study, the probability of DA2G congestion is $19.714\%$. This information is used to determine the congestion of DA2G base stations at each time-slot as input to the MILP models.

An important parameter in this study is cost, which has been considered in dollars per month. According to Amazon Web Services~\cite{vmtype}, Compute-Optimized instance types is suggested for high performance purposes, e.g., video streaming. Therefore, we assume all $k \in \mathcal{K}$ use the same \textit{c4.2xlarge} instance type, which costs $\$229$ per month~\cite{vmprice}. On the other hand, the link costs are associated to the bandwidth required by our services ($\sim20~Mbps$), which are $L_{G}=\$60$~\cite{groundcost} and $L_{sat}=\$130$~\cite{leocost}. Regarding the cost of $L_{DA2G}$, we assumed two values between $L_{G}$ and $L_{sat}$, $\$83$ and $\$107$. According to~\cite{wu2016energy}, service migration costs can be expressed as a function ($\mathcal{M}$) of bandwidth cost (i.e., $Cost_{uv}^L$) and instance memory size (i.e., $\propto Cost_k^{IN}$). Therefore, to simplify the model, we present a tunable instance migration cost as function $\mathcal{M}$:

{\small{\begin{align}
    Cost_k^{mig} = & \delta \times \mathcal{M}(Cost_{uv}^L, Cost_k^{IN}), (u,v) \in L_G, k \in \mathcal{K} \label{eq:delta}
\end{align}}}
\noindent where $0\leq\delta\leq1$ is the weighting factor. To simplify our model, we set the $\mathcal{M}$ as a sum of  $Cost_k^{IN}$ and $Cost_{uv}^L$. Nevertheless, more factors such as server CPU utilization and application sensitivity against migration can be considered to formulate the migration cost function~\cite{wu2016energy}.
\begin{figure*}[t]
    \centering
    \begin{subfigure}[t]{0.31\textwidth}
        \includegraphics[width=\linewidth]{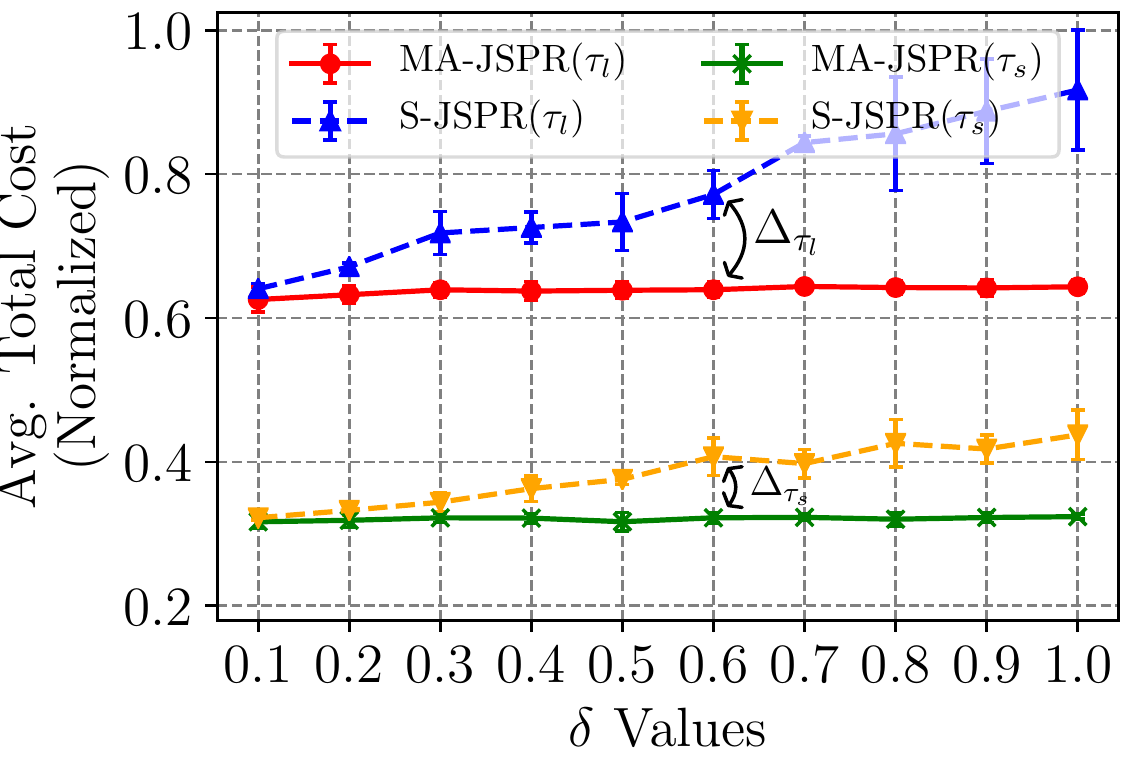}
        \caption{Average total cost}\label{fig:totalcost}
    \end{subfigure}~
    \begin{subfigure}[t]{0.31\textwidth}
        \includegraphics[width=\linewidth]{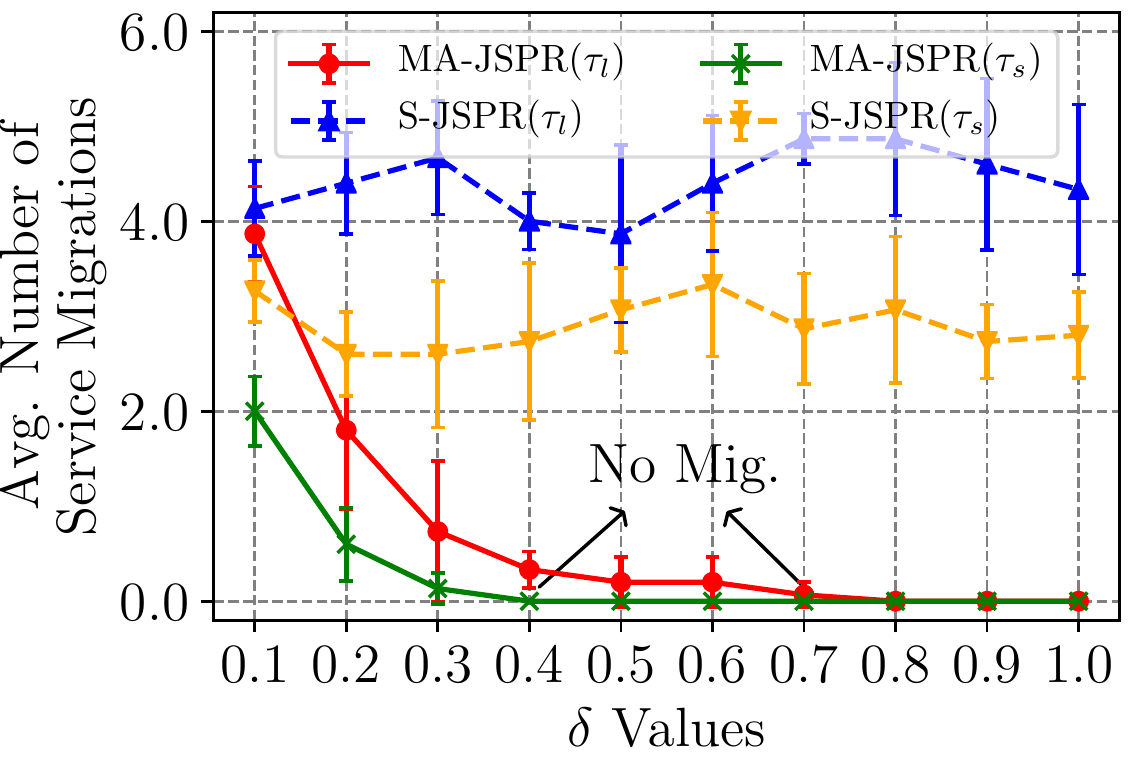}
        \caption{Average number of service migrations}\label{fig:migration}
    \end{subfigure}~
    \begin{subfigure}[t]{0.31\textwidth}
    \includegraphics[width=\linewidth]{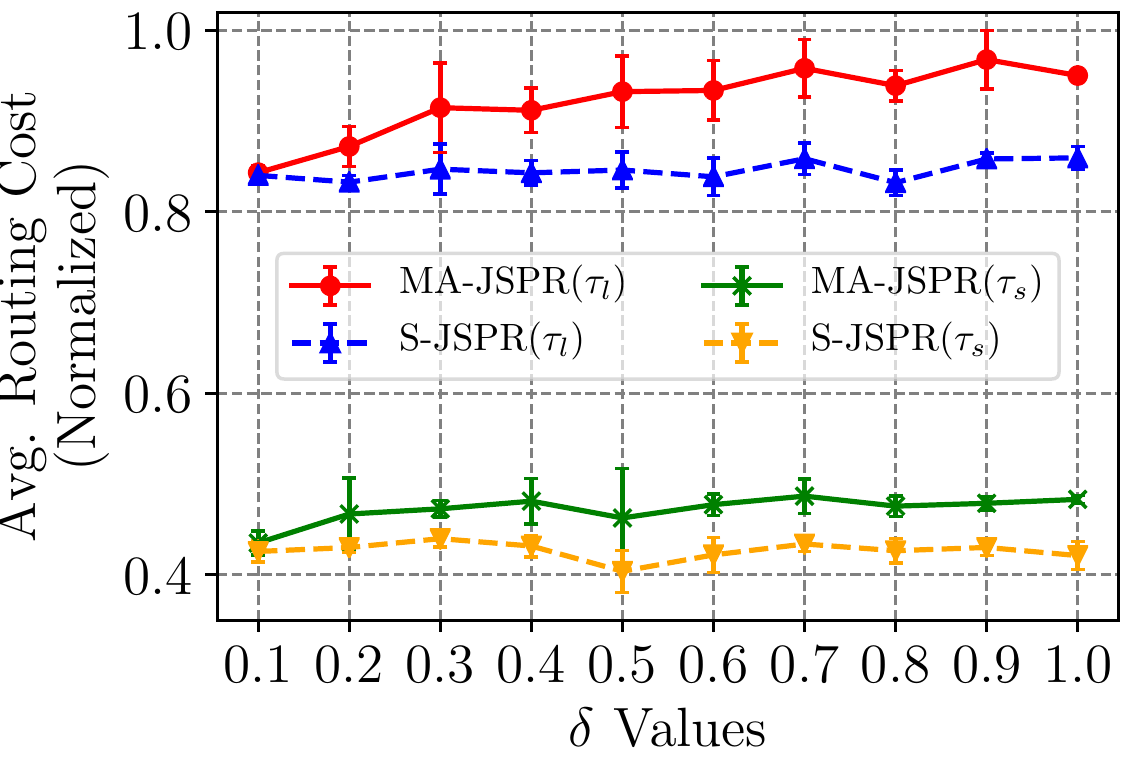}
    \caption{Average routing cost}\label{fig:routingcost}
\end{subfigure}
    \caption{MA-JSPR and S-JSPR comparison}\label{fig:comparison}
\end{figure*}

We modelled the SAGIN graph $\mathcal{G}$ by using Networkx~$2.1$ as the input of the MILP models. The proposed MILP models has been implemented using Gurobipy Optimizer~8.0.1 with $5\%$ solution gap for scalability purposes. The simulations were executed on a desktop computer, equipped with Intel Core i7-6700 @3.40~GHz CPU, 16~GB of RAM, running Ubuntu 18.04 x64 OS. We note that, due to lack of space, we omit the results for $L_{DA2G}$ cost of $\$107$. Nevertheless, we observe similar results by using this value. A detailed demonstration of the scenario presented in this work is available in~\cite{varasteh2019infocom}.

\textbf{\textit{B. Results:}} We compare MA-JSPR and S-JSPR for the above scenario. In order to evaluate the impact of different service migration costs with respect the network and service instance deployment costs (Eq.~(\ref{eq:delta})), we vary the $\delta$ value between $0.1$ and $1$, with step size of $0.1$. In our first study, we compare the total cost (including migration, network and service instance deployment costs) when applying MA-JSPR and S-JSPR for the two different types of flights: short flights with $\tau_s = 3$ and long flights with $\tau_l = 7$. The models have been evaluated for 30 random flights; results are batched and the mean and standard deviation is reported. Also, the impact of having other flights flying in the same area have been considered by the DA2G link congestion probability.

The total cost of the models is compared in Fig.~\ref{fig:totalcost}. It can the observed that the costs of MA-JSPR are always lower than S-JSPR and the difference increases with the increase of migration cost. Also, it can be seen that the cost of flight with longer duration ($\tau_l$) is higher than the short flight ($\tau_s$), because the amount of required resources is higher for the long flight. Also, the number of migrations is expected to be higher for the longer flights, which contributes to the total cost. Moreover, it can be seen that the cost difference for long flight is higher than the short flight ($\Delta_{\tau_l} > \Delta_{\tau_s}$). Hence, it can be concluded that the amount of cost savings is higher when more information about the future is taken into account. It has to be mentioned that in our use case, the future positions of the flights are known, which makes the problem simpler to solve. In other use cases, such as autonomous vehicles, the vehicle mobility pattern and its future positions rely on models, which are not easy to determine.    
\begin{figure}[b]
\centering
\includegraphics[width=.65\columnwidth]{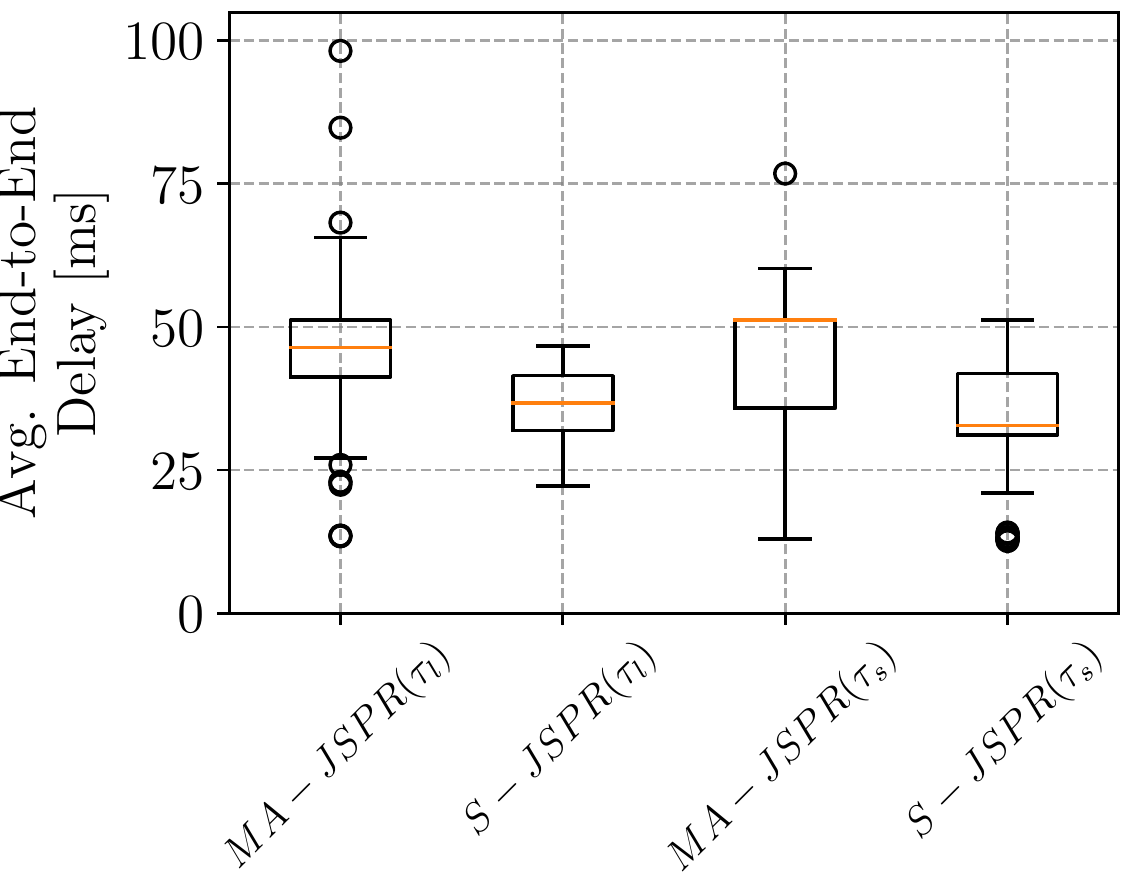}
\caption{Comparison of the average end-to-end delay}
\label{fig:delay}
\end{figure}

Secondly, Fig.~\ref{fig:migration} shows the comparison of the the number of service migrations for MA-JSPR and S-JSPR. It can be observed that the average number of service migrations in S-JSPR is significantly higher than MA-JSPR since it does not take the future flight locations into account. Moreover, it can be seen that the number of service migrations for the long flight $\tau_l$ is always higher than $\tau_s$, since it is flying for a longer distance and the probability of migration raises. The figure also shows that the number of service migrations decreases with the increase of migration cost ($\delta$). However, this can form a trade-off between the routing cost and migration cost (or number of service migrations). For example, if the migration cost is high and the flight is moving, in future flight positions more routing cost needs to be paid to reach to the service instance. This trade-off can be seen in Fig.~\ref{fig:routingcost}. As is reported, for the $\tau_s$ and $\tau_l$, MA-JSPR model triggers no service migrations at $\delta = 0.4$ and $\delta=0.7$, respectively (see Fig.~\ref{fig:migration}). Accordingly, as Fig.~\ref{fig:routingcost} presents, the routing cost in these cases increase from $\delta=0.1$ to $\delta=0.4$ and $\delta=0.7$. After $\delta=0.4$ and $\delta=0.7$, the routing cost does not increase and shows fluctuations, since no service migration is triggered anymore. 
\begin{table}[b]
\setlength\belowcaptionskip{12pt}
\caption{Average runtime for different MILP models [seconds]}
\label{table:runtime}
\tiny
\resizebox{\columnwidth}{!}{
\begin{tabular}{cccccccc}
\toprule
\textbf{$\delta$ Values} & \textbf{0.1} & \textbf{0.3} & \textbf{0.5} & \textbf{0.7} & \textbf{0.9} \\ \hline
\textbf{MA-JSPR($\tau_l$)} & 1997.45 & 1936.02 & 3310.87 & 2496.06& 1697.35 \\ \hline
\textbf{S-JSPR($\tau_l$)} & 108.67& 98.61 & 111.30 &
101.65 & 105.58 \\ \hline
\textbf{MA-JSPR($\tau_s$)} & 203.74 & 259.45 & 261.15 & 214.67 & 238.62 \\\hline
\textbf{S-JSPR($\tau_s$)} & 52.35 & 52.32 &  52.41 &
53.35 & 48.51 \\
\bottomrule
\end{tabular}}
\end{table}
Additionally, Fig.~\ref{fig:routingcost} shows that S-JSPR model produces lower routing costs compared to MA-JSPR because it neglects future flight positions and hence, it places the service instance at the closest DC (i.e., low routing cost). We note that, MA-JSPR can control the number of service migrations, while meeting the end-to-end required delay of each service request. 

Besides, the achieved average end-to-end delay of both services for S-JSPR and MA-JSPR is compared in Fig.~\ref{fig:delay}. It can be observed that the service requirements are generally met. Also, we see that the delay is mostly below $50~ms$, which means the DA2G is utilized most of the time to route the services to the selected DC. Moreover, Fig.~\ref{fig:delay} shows that in both $\tau_l$ and $\tau_s$ cases, S-JSPR can achieve lower average end-to-end delay than MA-JSPR. This is due to the fact that S-JSPR does not consider the future flight positions to solve the problem. In our scenario, there is a local DC node close to the satellite gateway nodes in the ground $N_{SG}$. Consequently, it can be seen that the average delay for MA-JSPR($\tau_s$) is higher than MA-JSPR($\tau_l$). This is because the short flight $\tau_s$ is passing by fewer DCs during the flight compared to the longer $\tau_l$ one. Therefore, MA-JSPR($\tau_s$) would prefer to use the local DC close to the satellite gateways to avoid the routing cost to reach farther DCs through the DA2G links. Additionally, it can be observed that the maximum achieved end-to-end delay in MA-JSPR($\tau_l$) is higher than MA-JSPR($\tau_s$). The reason is that in $\tau_l$ case, the paths would be longer than $\tau_s$ case to improve the cost-efficiency (see Fig.~\ref{fig:routingcost}). Notably, the average end-to-end delay can be affected significantly by the DA2G link congestion.

Finally, we present the runtime comparison of MA-JSPR and S-JSPR in Table~\ref{table:runtime}. It can be seen that the runtime for long flight is higher than for the short ones. Moreover, Table~\ref{table:runtime} indicates that the time complexity of S-JSPR is significantly lower than MA-JSPR, since it does not consider the future flight positions. 
\vspace{-.1cm}
\section{Conclusion}\label{sec:conclusion}
\vspace{-.04cm}
In this paper, two Mixed Integer Linear Programs (MILPs) have been presented for the Joint Service Placement and Routing~(JSPR) problem: \textit{i)} Static JSPR (S-JSPR), and \textit{ii)} Mobility-Aware JSPR (MA-JSPR) in Space-Air-Ground Integrated Networks~(SAGIN). The former considers one single time-slot, while the latter one considers a set of future flight positions for solving the JSPR problem and minimizing the total cost. To evaluate the proposed MILPs, we have considered a European-based SAGIN, which includes the satellite network, the Direct-Air-To-Ground~(DA2G) network, the ground network, as well as the flight and DC locations. We demonstrated MA-JSPR is able to utilize \ac{DA2G} and satellite connections to satisfy the passenger service requests in a cost-effective manner, compared to S-JSPR. In addition, we showed a trade-off between routing and migration costs. Also, we showed that a service migration model can avoid unnecessary migrations and improve the long-term cost-efficiency of MA-JSPR. Additionally, our evaluation results showed that S-JSPR achieves higher QoS levels and lower runtime compared to MA-JSPR. The models provided in this paper can help airlines to improve the network planning to support the services of their passengers. Yet, they can tune the inputs of the proposed models (e.g., cost values and service requirements) to fit the models to their particular use-case. Also, this paper can help them to compare the impact of different service providers (e.g., network/satellite providers) on their long-term total cost.
\vspace{-.1cm}
\section{Acknowledgment}
\vspace{-.05cm}
This work was supported in part under the Celtic-Plus subproject SEcure Networking for a DATacenter cloud in Europe (SENDATE)-PLANETS (Project ID 16KIS0261/16KIS0461) funded by the German Federal Ministry of Education and Research (BMBF). This work reflects only the authors’ view and the funding agency is not responsible for any use that may be made of the information it contains. The authors would like to thank Raphael Durner, Arled Papa, and the reviewers for their valuable comments.
\vspace{-.12cm}
\bibliographystyle{unsrt}
\bibliography{main}
\begin{acronym}[DA2G]
 \acro{A2G}{Air-to-Ground}
 \acro{DA2G}{Direct Air-To-Ground}
 \acro{GEO}{Geostationary Earth Orbit}
 \acro{LEO}{Low Earth Orbit}
 \acro{RTT}{Round Trip Time}
 \acro{QoS}{Quality of Service}
\end{acronym}

\end{document}